\documentclass[twocolumn,showpacs,preprintnumbers,amsmath,amssymb,superscriptaddress,nofootinbib]{revtex4}

\usepackage{graphicx}
\usepackage{dcolumn}
\usepackage{bm}
\usepackage[figuresright]{rotating}
\usepackage{fancyhdr}
\usepackage{enumerate}
\usepackage{indentfirst}
\usepackage{xspace}
\usepackage{color}

\begin{document}

\title{\bf Measurement of neutral current coherent $\pi^{0}$ production on carbon in a few-GeV neutrino beam}

\affiliation{Institut de Fisica d'Altes Energies, Universitat Autonoma de Barcelona, E-08193 Bellaterra (Barcelona), Spain}
\affiliation{Department of Physics, University of Colorado, Boulder, Colorado 80309, USA}
\affiliation{Department of Physics, Columbia University, New York, NY 10027, USA}
\affiliation{Fermi National Accelerator Laboratory; Batavia, IL 60510, USA}
\affiliation{High Energy Accelerator Research Organization (KEK), Tsukuba, Ibaraki 305-0801, Japan}
\affiliation{Department of Physics, Imperial College London, London SW7 2AZ, UK}
\affiliation{Department of Physics, Indiana University, Bloomington, IN 47405, USA}
\affiliation{Kamioka Observatory, Institute for Cosmic Ray Research, University of Tokyo, Gifu 506-1205, Japan}
\affiliation{Research Center for Cosmic Neutrinos, Institute for Cosmic Ray Research, University of Tokyo, Kashiwa, Chiba 277-8582, Japan}
\affiliation{Department of Physics, Kyoto University, Kyoto 606-8502, Japan}
\affiliation{Los Alamos National Laboratory; Los Alamos, NM 87545, USA}
\affiliation{Department of Physics and Astronomy, Louisiana State University, Baton Rouge, LA 70803, USA}
\affiliation{Department of Physics, Massachusetts Institute of Technology, Cambridge, MA 02139, USA}
\affiliation{Department of Chemistry and Physics, Purdue University Calumet, Hammond, IN 46323, USA}
\affiliation{Universit$\grave{a}$ di Roma La Sapienza, Dipartimento di Fisica and INFN, I-00185 Rome, Italy}
\affiliation{Physics Department, Saint Mary's University of Minnesota, Winona, MN 55987, USA}
\affiliation{Department of Physics, Tokyo Institute of Technology, Tokyo 152-8551, Japan}
\affiliation{Instituto de Fisica Corpuscular, Universidad de Valencia and CSIC, E-46071 Valencia, Spain}

\author{Y.~Kurimoto}\affiliation{High Energy Accelerator Research Organization (KEK), Tsukuba, Ibaraki 305-0801, Japan}
\author{J.~L.~Alcaraz-Aunion}\affiliation{Institut de Fisica d'Altes Energies, Universitat Autonoma de Barcelona, E-08193 Bellaterra (Barcelona), Spain}
\author{S.~J.~Brice}\affiliation{Fermi National Accelerator Laboratory; Batavia, IL 60510, USA}
\author{L.~Bugel}\affiliation{Department of Physics, Massachusetts Institute of Technology, Cambridge, MA 02139, USA}
\author{J.~Catala-Perez}\affiliation{Instituto de Fisica Corpuscular, Universidad de Valencia and CSIC, E-46071 Valencia, Spain}
\author{G.~Cheng}\affiliation{Department of Physics, Columbia University, New York, NY 10027, USA}
\author{J.~M.~Conrad}\affiliation{Department of Physics, Massachusetts Institute of Technology, Cambridge, MA 02139, USA}
\author{Z.~Djurcic}\affiliation{Department of Physics, Columbia University, New York, NY 10027, USA}
\author{U.~Dore}\affiliation{Universit$\grave{a}$ di Roma La Sapienza, Dipartimento di Fisica and INFN, I-00185 Rome, Italy}
\author{D.~A.~Finley}\affiliation{Fermi National Accelerator Laboratory; Batavia, IL 60510, USA}
\author{A.~J.~Franke}\affiliation{Department of Physics, Columbia University, New York, NY 10027, USA}
\author{C.~Giganti}
\altaffiliation[Present address: ]{DSM/Irfu/SPP, CEA Saclay, F-91191 Gif-sur-Yvette, France}
\affiliation{Universit$\grave{a}$ di Roma La Sapienza, Dipartimento di Fisica and INFN, I-00185 Rome, Italy}
\author{J.~J.~Gomez-Cadenas}\affiliation{Instituto de Fisica Corpuscular, Universidad de Valencia and CSIC, E-46071 Valencia, Spain}
\author{P.~Guzowski}\affiliation{Department of Physics, Imperial College London, London SW7 2AZ, UK}
\author{A.~Hanson}\affiliation{Department of Physics, Indiana University, Bloomington, IN 47405, USA}
\author{Y.~Hayato}\affiliation{Kamioka Observatory, Institute for Cosmic Ray Research, University of Tokyo, Gifu 506-1205, Japan}
\author{K.~Hiraide}
\altaffiliation[Present address: ]{Kamioka Observatory, Institute for Cosmic Ray Research, University of Tokyo, Gifu 506-1205, Japan}
\affiliation{Department of Physics, Kyoto University, Kyoto 606-8502, Japan}
\author{G.~Jover-Manas}\affiliation{Institut de Fisica d'Altes Energies, Universitat Autonoma de Barcelona, E-08193 Bellaterra (Barcelona), Spain}
\author{G.~Karagiorgi}\affiliation{Department of Physics, Massachusetts Institute of Technology, Cambridge, MA 02139, USA}
\author{T.~Katori}\affiliation{Department of Physics, Indiana University, Bloomington, IN 47405, USA}
\author{Y.~K.~Kobayashi}\affiliation{Department of Physics, Tokyo Institute of Technology, Tokyo 152-8551, Japan}
\author{T.~Kobilarcik}\affiliation{Fermi National Accelerator Laboratory; Batavia, IL 60510, USA}
\author{H.~Kubo}\affiliation{Department of Physics, Kyoto University, Kyoto 606-8502, Japan}
\author{W.~C.~Louis}\affiliation{Los Alamos National Laboratory; Los Alamos, NM 87545, USA}
\author{P.~F.~Loverre}\affiliation{Universit$\grave{a}$ di Roma La Sapienza, Dipartimento di Fisica and INFN, I-00185 Rome, Italy}
\author{L.~Ludovici}\affiliation{Universit$\grave{a}$ di Roma La Sapienza, Dipartimento di Fisica and INFN, I-00185 Rome, Italy}
\author{K.~B.~M.~Mahn}
\altaffiliation[Present address: ]{TRIUMF, Vancouver, British Columbia, V6T 2A3, Canada}
\affiliation{Department of Physics, Columbia University, New York, NY 10027, USA}
\author{C.~Mariani}\affiliation{Department of Physics, Columbia University, New York, NY 10027, USA}
\author{S.~Masuike}\affiliation{Department of Physics, Tokyo Institute of Technology, Tokyo 152-8551, Japan}
\author{K.~Matsuoka}\affiliation{Department of Physics, Kyoto University, Kyoto 606-8502, Japan}
\author{V.~T.~McGary}\affiliation{Department of Physics, Massachusetts Institute of Technology, Cambridge, MA 02139, USA}
\author{W.~Metcalf}\affiliation{Department of Physics and Astronomy, Louisiana State University, Baton Rouge, LA 70803, USA}
\author{G.~B.~Mills}\affiliation{Los Alamos National Laboratory; Los Alamos, NM 87545, USA}
\author{G.~Mitsuka}
\altaffiliation[Present address: ]{Solar-Terrestrial Environment Laboratory, Nagoya University, Furo-cho, Chikusa-ku, Nagoya, Japan}
\affiliation{Research Center for Cosmic Neutrinos, Institute for Cosmic Ray Research, University of Tokyo, Kashiwa, Chiba 277-8582, Japan}
\author{Y.~Miyachi}
\altaffiliation[Present address: ]{Department of Physics, Yamagata University, Yamagata, 990-8560 Japan}
\affiliation{Department of Physics, Tokyo Institute of Technology, Tokyo 152-8551, Japan}
\author{S.~Mizugashira}\affiliation{Department of Physics, Tokyo Institute of Technology, Tokyo 152-8551, Japan}
\author{C.~D.~Moore}\affiliation{Fermi National Accelerator Laboratory; Batavia, IL 60510, USA}
\author{Y.~Nakajima}\affiliation{Department of Physics, Kyoto University, Kyoto 606-8502, Japan}
\author{T.~Nakaya}\affiliation{Department of Physics, Kyoto University, Kyoto 606-8502, Japan}
\author{R.~Napora}\affiliation{Department of Chemistry and Physics, Purdue University Calumet, Hammond, IN 46323, USA}
\author{P.~Nienaber}\affiliation{Physics Department, Saint Mary's University of Minnesota, Winona, MN 55987, USA}
\author{D.~Orme}\affiliation{Department of Physics, Kyoto University, Kyoto 606-8502, Japan}
\author{M.~Otani}\affiliation{Department of Physics, Kyoto University, Kyoto 606-8502, Japan}
\author{A.~D.~Russell}\affiliation{Fermi National Accelerator Laboratory; Batavia, IL 60510, USA}
\author{F.~Sanchez}\affiliation{Institut de Fisica d'Altes Energies, Universitat Autonoma de Barcelona, E-08193 Bellaterra (Barcelona), Spain}
\author{M.~H.~Shaevitz}\affiliation{Department of Physics, Columbia University, New York, NY 10027, USA}
\author{T.-A.~Shibata}\affiliation{Department of Physics, Tokyo Institute of Technology, Tokyo 152-8551, Japan}
\author{M.~Sorel}\affiliation{Instituto de Fisica Corpuscular, Universidad de Valencia and CSIC, E-46071 Valencia, Spain}
\author{R.~J.~Stefanski}\affiliation{Fermi National Accelerator Laboratory; Batavia, IL 60510, USA}
\author{H.~Takei}
\altaffiliation[Present address: ]{Kitasato University, Tokyo, 108-8641 Japan}
\affiliation{Department of Physics, Tokyo Institute of Technology, Tokyo 152-8551, Japan}
\author{H.-K.~Tanaka}\affiliation{Department of Physics, Massachusetts Institute of Technology, Cambridge, MA 02139, USA}
\author{M.~Tanaka}\affiliation{High Energy Accelerator Research Organization (KEK), Tsukuba, Ibaraki 305-0801, Japan}
\author{R.~Tayloe}\affiliation{Department of Physics, Indiana University, Bloomington, IN 47405, USA}
\author{I.~J.~Taylor}
\altaffiliation[Present address: ]{Department of Physics and Astronomy, State University of New York, Stony Brook, NY 11794-3800, USA}
\affiliation{Department of Physics, Imperial College London, London SW7 2AZ, UK}
\author{R.~J.~Tesarek}\affiliation{Fermi National Accelerator Laboratory; Batavia, IL 60510, USA}
\author{Y.~Uchida}\affiliation{Department of Physics, Imperial College London, London SW7 2AZ, UK}
\author{R.~Van~de~Water}\affiliation{Los Alamos National Laboratory; Los Alamos, NM 87545, USA}
\author{J.~J.~Walding}\affiliation{Department of Physics, Imperial College London, London SW7 2AZ, UK}
\author{M.~O.~Wascko}\affiliation{Department of Physics, Imperial College London, London SW7 2AZ, UK}
\author{H.~B.~White}\affiliation{Fermi National Accelerator Laboratory; Batavia, IL 60510, USA}
\author{M.~J.~Wilking}
\altaffiliation[Present address: ]{TRIUMF, Vancouver, British Columbia, V6T 2A3, Canada}
\affiliation{Department of Physics, University of Colorado, Boulder, Colorado 80309, USA}
\author{M.~Yokoyama}
\altaffiliation[Present address: ]{Department of Physics, University of Tokyo, Tokyo 113-0033, Japan}
\affiliation{Department of Physics, Kyoto University, Kyoto 606-8502, Japan}
\author{G.~P.~Zeller}\affiliation{Los Alamos National Laboratory; Los Alamos, NM 87545, USA}
\author{E.~D.~Zimmerman}\affiliation{Department of Physics, University of Colorado, Boulder, Colorado 80309, USA}

\collaboration{The SciBooNE Collaboration}\noaffiliation 

\begin{abstract}

The SciBooNE Collaboration reports a measurement of neutral current
coherent $\pi^{0}$ production on carbon by a muon neutrino beam with
average energy 0.8 GeV. The separation of coherent from inclusive
$\pi^{0}$ production has been improved by detecting recoil protons
from resonant $\pi^{0}$ production.  We measure the ratio of the
neutral current coherent $\pi^{0}$ production to total charged current
cross sections to be $(1.16~\pm~0.24) \times 10^{-2}$.  The ratio of
charged current coherent $\pi^+$ to neutral current coherent $\pi^{0}$
production is calculated to be $0.14^{+0.30}_{-0.28}$, using our
published charged current coherent pion measurement.
\end{abstract}
\date{\today}

\maketitle
\newcommand{\pzero}{$\pi^{\rm{0}}$}
\section{Introduction} 
\label{sec:introduction}
Recent measurements of coherent pion production by muon neutrinos at
neutrino energies around 1 GeV have inspired significant
discussion~\cite{Morfin:2009zz}.  In coherent pion production, the
neutrino interacts with an entire nucleus; no nucleon recoil
occurs and the {\pzero} tends to be emitted in the forward direction.

For charged current (CC) coherent pion production, both K2K and
SciBooNE set limits on the ratio of CC coherent pion production to the
total CC cross sections near
1~GeV\cite{Hasegawa:2005td,Hiraide:2008eu}, These published upper
limits are significantly lower than those predicted by the Rein and
Sehgal model~\cite{Rein:1982pf,Rein:2006di} which is widely used for
many neutrino oscillation experiments.  Meanwhile, evidence for
neutral current (NC) coherent pion production with neutrino energy
less than 2 GeV has been reported by the MiniBooNE Collaboration
\cite{AguilarArevalo:2008xs}.  The SciBooNE collaboration also
reported nonzero NC coherent pion production \cite{Kurimoto:2009wq}
although the result is only 1.6 standard deviations above zero
coherent production. Currently there is no theoretical model which can
accommodate all of these recent measurements.  Further experimental
inputs may help the development of theoretical models.

NC coherent pion production at neutrino energies around 1 GeV is also
important for neutrino oscillation experiments as a substantial
contribution to NC \pzero\ production (NC\pzero). The largest
contribution to NC\pzero\ is NC resonant pion production, in which the
neutrino interacts with a single nucleon in the target nucleus and
excites it to a baryon resonance; the resonant decay produces a pion
and a nucleon.  NC \pzero\ production is the largest $\nu_\mu$-induced
background in neutrino experiments searching for
$\nu_\mu\rightarrow\nu_e$ oscillations.  NC\pzero\ events cannot be
distinguished from $\nu_e$ signal events when, for example, one of the
two photons associated with $\pi^{0}\rightarrow\gamma\gamma$ is not
detected.

Both MiniBooNE's and SciBooNE's previous measurements of NC coherent
pion production were performed using only emitted \pzero\ kinematics.
However, in addition to the \pzero\ kinematics, the absence of a recoil
nucleon is a clear and less model-dependent feature of coherent pion
production. In SciBooNE, detection of the recoil nucleon is possible
using the fully active and fine-grained vertex detector, SciBar.

In this paper, we report a measurement of NC coherent \pzero\
production using a new analysis method in which the lack of recoil
nucleons is used to extract the fraction of coherent pions
within the inclusive \pzero\ dataset.  SciBooNE's full neutrino data
set, corresponding to $0.99 \times 10^{20}$ protons on target, is
used. To simulate coherent $\pi$ production, the Rein and Sehgal
model~\cite{Rein:1982pf}, including lepton mass
corrections~\cite{Rein:2006di}, is used. The axial vector mass $M_{A}$
and the nuclear radius parameter $R_{0}$ used in the model are set to
1.0 ${\rm GeV}/c^2$ and 1.0 fm, respectively. These are the same
values used in previous SciBooNE
papers~\cite{Kurimoto:2009wq,Hiraide:2008eu}.  This paper updates
our previous result\cite{Kurimoto:2009wq}, so, not only the
coherent $\pi$ production model but all simulations and the
experimental setup used in this analysis are the same as previously
described.

\section{NC \pzero\ event selections}
\label{sec:evselect}

The SciBooNE detector is comprised of three subsystems: a
scintillating bar neutrino vertex detector called SciBar, an
electromagnetic calorimeter, and a muon range detector.  We use SciBar
as the neutrino target as well as the particle tracker for this
analysis.  SciBar consists of 14336 polystyrene (${\rm C}_{8}{\rm
H}_{8}$) scintillator bars.  The scintillators are arranged vertically
and horizontally to construct a $3\times3\times1.7~{\rm m}^{3}$ volume
with a total mass of 15 tons.

\begin{figure}[bp]
  \begin{center}
    \includegraphics*[scale=0.85]{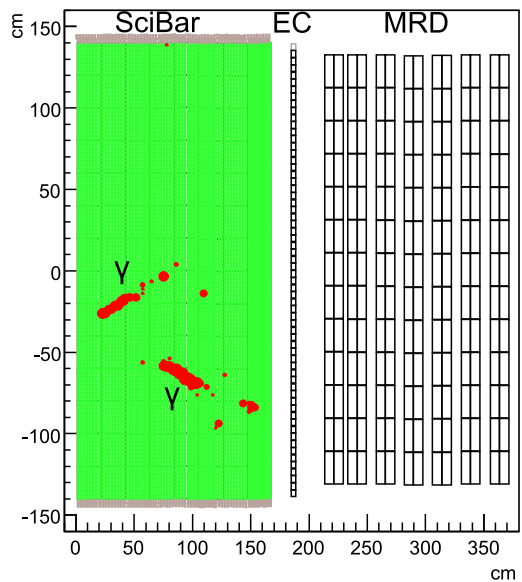}
    \caption{Event display of a typical NC\pzero\ event candidate in SciBooNE data.  
    The neutrino beam runs from left to right in this figure,
    encountering SciBar, the electromagnetic calorimeter (EC) and
    the muon range detector (MRD), in that order. The circles on SciBar
    indicate ADC hits for which the
    area of the circle is proportional to the energy deposition 
    in that channel. This event display shows the electromagnetic 
    shower tracks from the pair conversions of the two \pzero\ 
    decay photons. }
    \label{fig:scibar_schematic}
  \end{center}
\end{figure}

As shown in Fig.~\ref{fig:scibar_schematic}, 
NC \pzero\ production is observed as two isolated tracks in SciBar
due to two gamma rays, coming from the decay of the \pzero, converted
into two $e^+e^-$ pairs. 
The background events stem from sources both
internal and external to SciBar. Internal backgrounds are neutrino
interactions other than NC\pzero\ (mainly CC) within SciBar. 
External backgrounds come from neutrino interactions in the material 
outside of the detector volume (dirt background events) as well as 
cosmic rays. To reduce these background events, several event 
selections are performed before extracting coherent {\pzero}s. 

To reject CC background events, identifying muons is paramount.  We
identify the following three types of tracks as muons and reject such
events, (i) tracks escaping from the side of SciBar, (ii) tracks
stopping in SciBar with delayed timing hits due to the decay electrons
and (iii) tracks penetrating the electromagnetic calorimeter located
at the downstream of SciBar. For the dirt background rejection, we use
the upstream part of SciBar as a charged particle veto and require
that the reconstructed vertices of both gamma ray candidates be in
SciBar. Finally, we require that the reconstructed invariant mass of
two gamma ray candidates be close to the \pzero\ mass.  All selections
are identical to those used in the previous
analysis~\cite{Kurimoto:2009wq} and are described in detail there.

After event selection, 657 events remain.  Subtracting the estimated
background of 240 events (202 internal and 38 external) yields 417
signal events.  The MC expectation is 368 events. The numbers and
distributions obtained by the MC simulation are normalized with the CC
data sample \cite{Kurimoto:2009wq}.  The purity of NC \pzero\
production after all event selections is estimated to be 61\%.  The
efficiency for NC\pzero\ production is estimated to be 5.3\%. The
efficiency for NC coherent \pzero production, incoherent \pzero\
production\footnote{NC incoherent \pzero\ production is defined 
as all NC\pzero\ events except for coherent \pzero\ production.   
After event selections, 89\% of the incoherent events come from resonant pion production 
and the rest come from deep inelastic 
scattering.} with recoil neutron and 
with recoil proton are estimated to
be 7.6\%, 6.2\% and 4.5\%\footnote{High track multiplicity around the
neutrino interaction vertex due to the proton recoil can cause
mis-reconstruction of the event.}.

\section{Coherent \pzero\ event selection}
\label{sec:cohfrac}

\begin{figure}[htbp]
  \begin{center}
    \includegraphics[keepaspectratio=false,width=70mm]{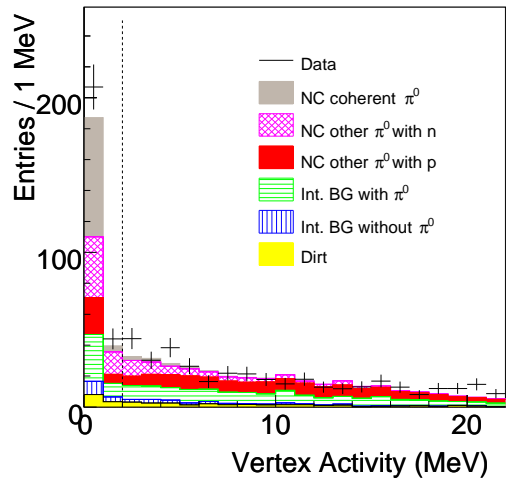}
  \end{center}
  \caption{Vertex activity after all event selections: 
    the contribution from NC coherent \pzero, incoherent NC\pzero\ with recoil neutrons, 
    incoherent  NC\pzero\ with recoil protons, 
    internal backgrounds with a {\pzero} in the final state, internal
    background without a {\pzero} in the final state and 
    ``dirt'' background events are shown separately for the MC
    simulation.}
  \label{fig:vtxact}
\end{figure}

In NC coherent pion production, there is no recoil nucleon in the
final state since the \pzero\ is produced by the neutrino interacting
with the whole nucleus.  Conversely, a recoiling nucleon should be
present in a resonant pion event. To separate the NC coherent \pzero\
events from the NC resonant \pzero\ events, recoil protons in the
final state are used. The recoil protons are detected by their large
energy deposition near the neutrino interaction vertex, so-called
vertex activity.  We search for the maximum deposited energy in a
scintillator strip around the reconstructed vertex, an area of
$40~{\rm cm}~\times~40~{\rm cm}$ in each view.  The choice of $40~{\rm
cm}$ ($\pm 20~{\rm cm}$ from the reconstructed vertex) for the area is
based on the vertex resolution which is approximately 12~cm for each
direction ($x$,$y$ and $z$). A \pzero\ at typical SciBooNE energies
travels, on average, $\sim$20~nm before decaying, so the reconstructed
intersection of the gamma tracks is a good estimate of the neutrino
interaction vertex.  Figure~\ref{fig:vtxact} shows the maximum
deposited energy distribution after all selections. Most of the
coherent \pzero\ contribution is peaked at zero while the other
\pzero\ events have high energy activity due to recoil protons. Events
with energy deposition greater than 2 MeV are considered to have
activity at the vertex.  Note that incoherent pion production
with a neutron recoil leaves no vertex activity unless the neutron
kicks off protons in the region where we search for the energy deposit.
Based on our MC simulation, the fraction of proton recoils in all incoherent
\pzero\ events is reduced from 71\% in the sample with vertex activity
 to 35\% in the sample without vertex activity.

\section{Data Analysis}
When a neutrino interacts with the entire nucleus, 
the following relation should be satisfied:
\begin{eqnarray}
  \label{eq:cohshiki}
  \frac{1}{|t|} > R,
\end{eqnarray}
where $t$ and $R$ are the four-momentum transfer to the target nucleus
from the neutrino and the radius of target nucleus, respectively.
This means that the cross section decreases rapidly when $1/|t|$
become smaller than $R$.  Using Eq.~\ref{eq:cohshiki}, we can deduce
\begin{eqnarray}
  \label{eq:cohshiki2}
  E_{\pi^{0}}(1-\cos\theta_{\pi^{0}}) < \frac{1}{R} \sim {\rm ~100~MeV},
\end{eqnarray}
following Ref.~\cite{Lackner:1979ax}. In this equation, $E_{\pi^{0}}$
and $\theta_{\pi^{0}}$ are the \pzero\ energy and direction with
respect to the neutrino beam, respectively.  From this fact, we can
determine the fraction of coherent \pzero\ production using the
reconstructed \pzero\ kinematic variable $E^{\rm
rec}_{{\pi}^{0}}(1-\cos{\theta}^{\rm rec}_{\pi^{\rm 0}})$, where
$E^{\rm rec}_{{\pi}^{0}}$ is the reconstructed \pzero\ energy
calculated as the sum of the reconstructed energies of two gamma ray
candidates and $\theta^{\rm rec}_{\pi^{\rm 0}}$ is the reconstructed
\pzero\ direction with respect to the neutrino beam axis.

\begin{figure}[htbp]
  \begin{center}
    \includegraphics[keepaspectratio=false,width=70mm]{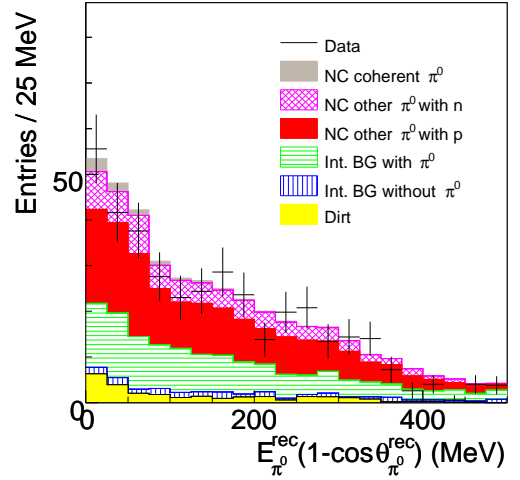}
    \includegraphics[keepaspectratio=false,width=70mm]{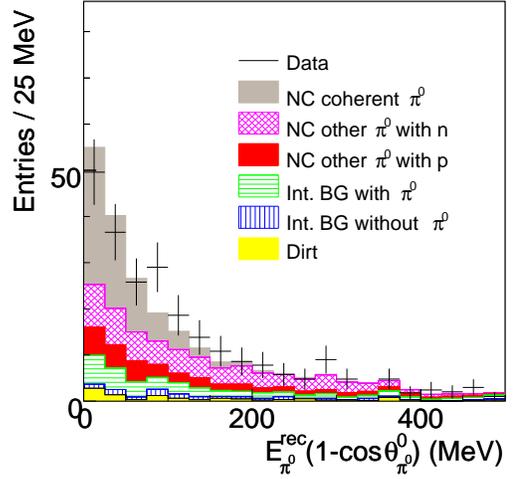}
  \end{center}
  \caption{ The $E^{\rm rec}_{{\pi}^{0}}(1-\cos{\theta}^{\rm
      rec}_{\pi^{\rm 0}})$ distributions after fitting with (top) and
    without (bottom) vertex activity.}
  \label{fig:pi0cohangleafterfit}
\end{figure}

\begin{figure}[htbp]
  \begin{center}
    \includegraphics[keepaspectratio=false,width=70mm]{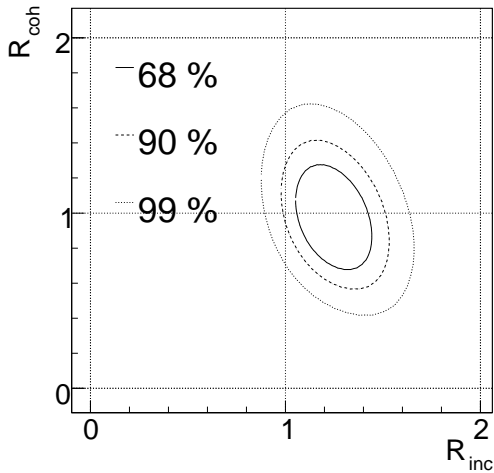}
  \end{center}
  \caption{The contours corresponding to 68\%, 90\% and 99\% confidence
    level for the fitted values of the scaling parameters; the number of
    degrees of freedom is 2.}
  \label{fig:contour}
\end{figure}

\begin{figure}[htbp]
  \begin{center}
    \includegraphics[keepaspectratio=false,width=70mm]{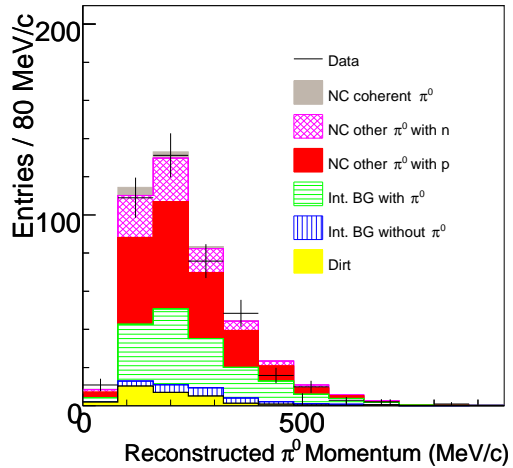}
    \includegraphics[keepaspectratio=false,width=70mm]{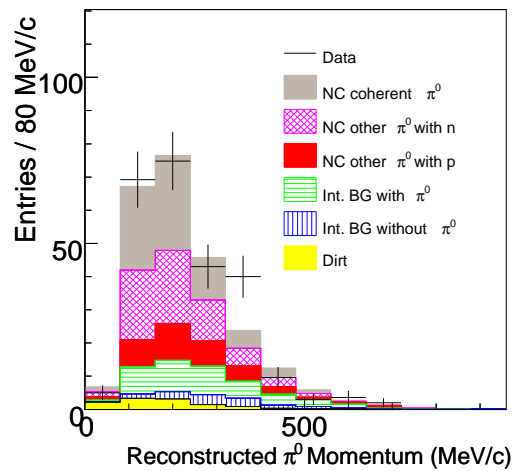}
  \end{center}
  \caption{Reconstructed \pzero\ momentum distributions after 
    fitting with the vertex activity (top) and without vertex
    activity (bottom).}
  \label{fig:pi0mom34_afterfit}
\end{figure}

\begin{figure}[htbp]
  \begin{center}
    \includegraphics[keepaspectratio=false,width=70mm]{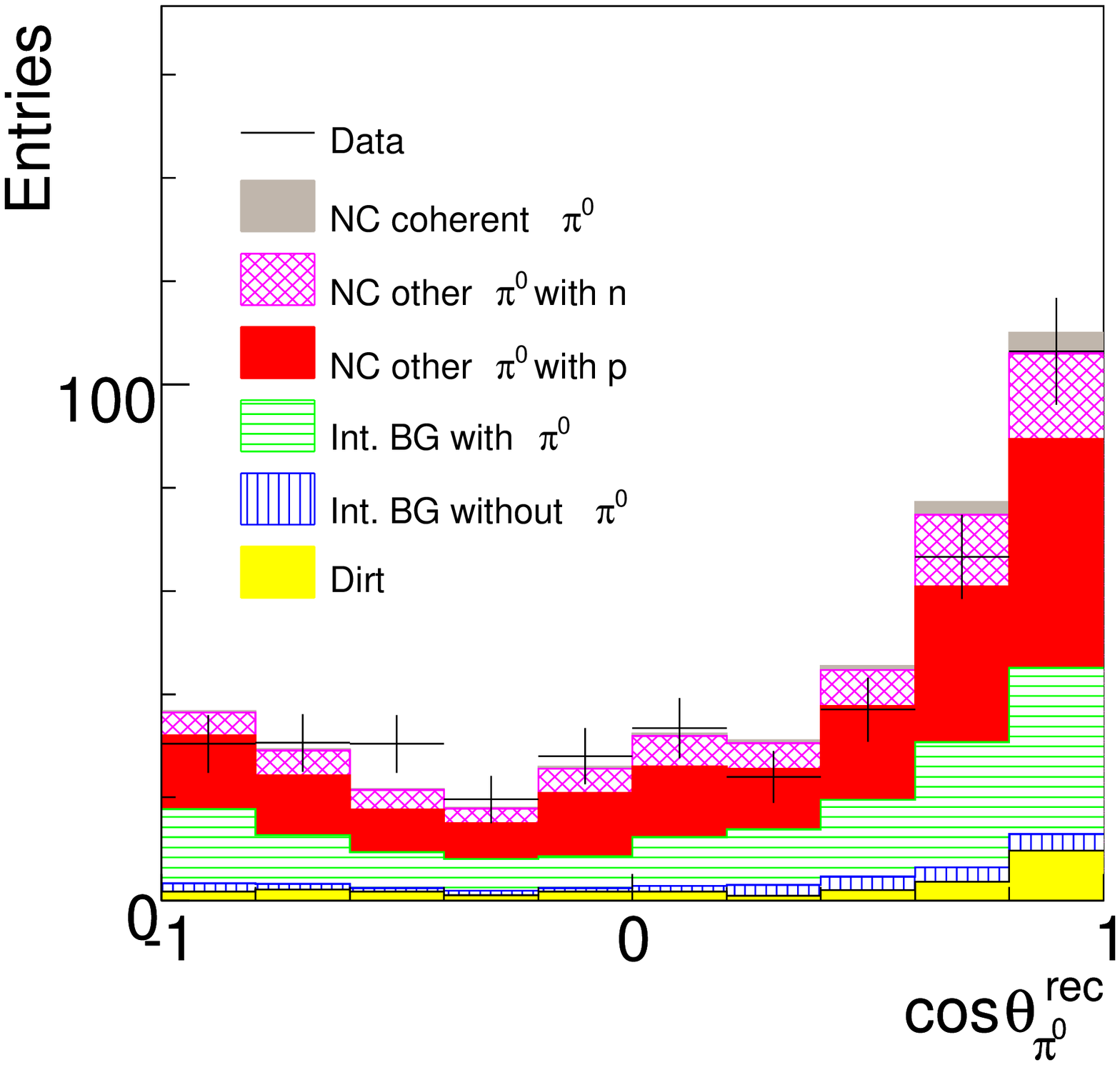}
    \includegraphics[keepaspectratio=false,width=70mm]{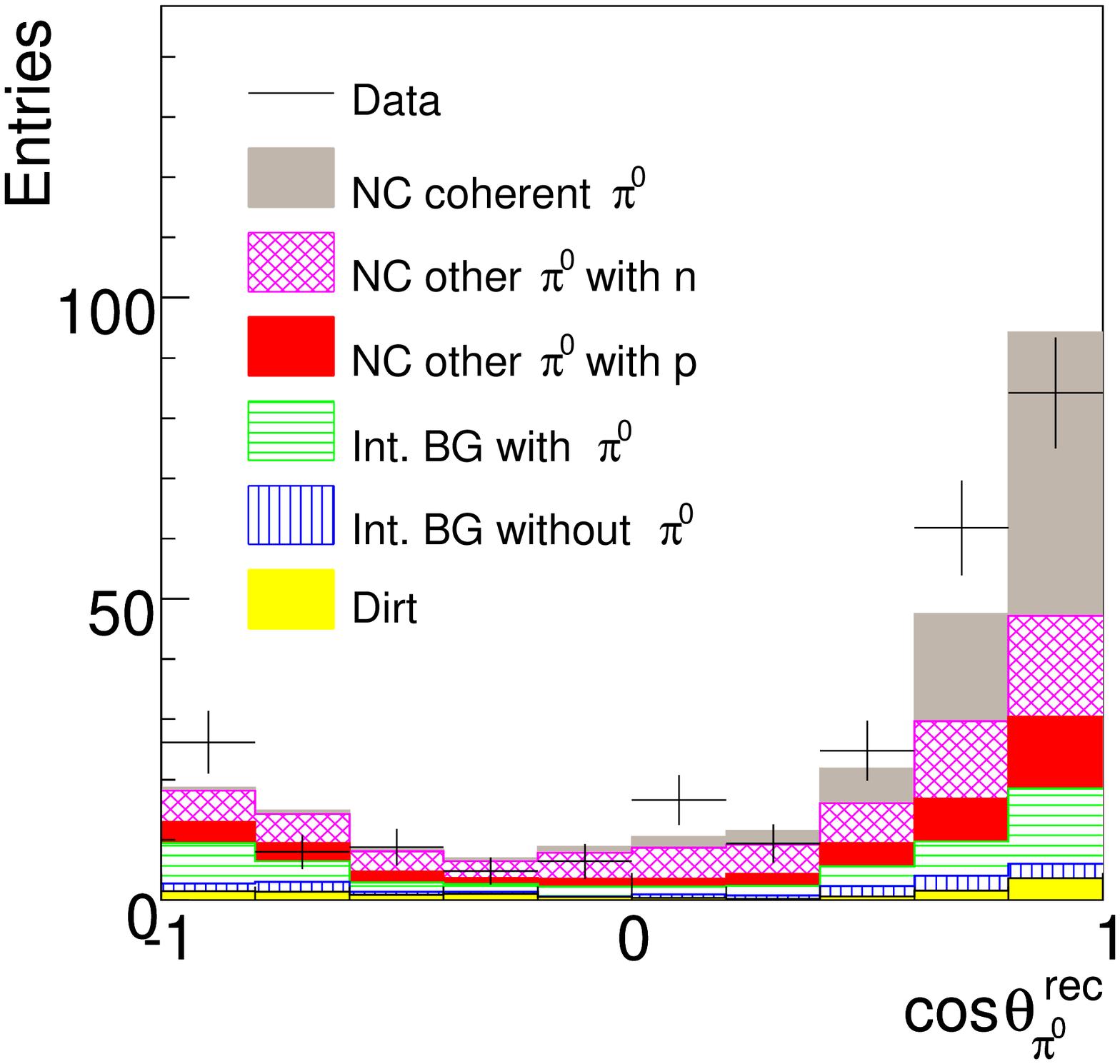}
  \end{center}
  \caption{ The $\cos\theta^{rec}_{\pi^{\rm 0}}$ distributions after 
    fitting with the vertex activity (top) and without vertex
    activity (bottom).}
  \label{fig:pi0angle_cos34_afterfit}
\end{figure}

We simultaneously fit two $E^{\rm rec}_{{\pi}^{0}}(1-\cos{\theta}^{\rm
rec}_{\pi^{\rm 0}})$ distributions, with and without the vertex
activity, with three templates made by dividing the final MC sample
into NC coherent \pzero, NC resonant \pzero\ and background samples.
Two parameters, ${\rm R}_{\rm coh}$ and ${\rm R}_{\rm inc}$ scale the
NC coherent \pzero\ and NC incoherent \pzero\ templates independently.
The background sample is fixed to the value of the MC prediction
although the systematic errors on the background prediction are taken
into account.  The expected number of events in the $i$-th bin in the
$E^{\rm rec}_{{\pi}^{0}}(1-\cos{\theta}^{\rm rec}_{\pi^{\rm 0}})$
distribution is expressed as:
\begin{eqnarray}
  N^{\rm exp}_{i} = {\rm R}_{\rm coh}\times N^{\rm coh}_{i} 
  +  {\rm R}_{\rm inc}\times N^{\rm inc}_{i}
  +  N^{\rm BG}_{i}.
\end{eqnarray}  
The fit minimizes the expression:
\begin{eqnarray}
  {\chi}^2 &=& -2\ln{\frac{f(N^{\rm obs};N^{\rm exp})}{f(N^{\rm obs};N^{\rm obs})}},
\end{eqnarray}
where $N^{\rm obs(exp)}$ represents the observed (expected) number of
events in all bins $(N^{\rm obs(exp)}_{1},N^{\rm
obs(exp)}_{2},{\ldots},N^{\rm obs(exp)}_{N})$ and $f(N^{\rm
obs};N^{\rm exp})$ is the Poisson likelihood to find $N^{\rm obs}$
events when $N^{\rm exp}$ events are expected.  When the systematic
errors for each bin and their correlation expressed with covariance
matrix $V_{jk}$ ($j,k=1,2,\ldots ,N(=39)$)\footnote {The total number
of bins for the two distributions is 40 and there is one bin without
entries. We do not include the empty bin in the fit.} are given, the
likelihood is expressed as
\begin{eqnarray}
  && f(N^{\rm obs};N^{\rm exp};V)
  = A\int_{}^{} \biggl[\Bigl[\prod^{N}_{i=1} 
      dx_{i}\frac{{x_{i}}^{N^{\rm obs}_{i}}e^{-{x_{i}}}}{N^{\rm obs}_{i}!}\Bigr]\nonumber \\
    && \times \exp\Bigl[-\frac{1}{2}\sum_{j=1}^{N}\sum_{k=1}^{N}(x_{j}-N^{\rm exp}_{j})V^{-1}_{jk}(x_{k}-N^{\rm exp}_{k})\Bigr]\biggr],
\end{eqnarray} 
where $A$ is a normalization constant. The details of the systematic
errors and the calculation of the integral are described in
Ref.~\cite{Kurimoto:2009wq}.
The result of the fit is:
\begin{eqnarray}
  \label{eqn:fitresult}
	{\rm R}_{\rm coh} &=& 0.96 \pm 0.20, \\
	{\rm R}_{\rm inc} &=& 1.24 \pm 0.13 .
\end{eqnarray}  
The $E^{\rm rec}_{{\pi}^{0}}(1-\cos{\theta}^{\rm rec}_{\pi^{\rm 0}})$
distribution after the fitting is shown in
Figure~\ref{fig:pi0cohangleafterfit}.  The $\chi^2$ per degree of
freedom (DOF), before the fit is 30.8/39 = 0.79, and it is 26.6/37 =
0.72 after the fit.  Figure~\ref{fig:contour} shows three contours
corresponding to 68\%, 90\% and 99\% confidence level. The statistical
error and all systematic errors are included in the errors of ${\rm
R}_{\rm coh}$ and ${\rm R}_{\rm inc}$.  Without the systematic errors,
we obtain $0.98 \pm 0.18({\rm stat.})$ and $1.19 \pm 0.10({\rm
stat.})$ for ${\rm R}_{\rm coh}$ and ${\rm R}_{\rm inc}$,
respectively. Hence, the uncertainty of the measurement is dominated
by the statistical uncertainty.  Figures~\ref{fig:pi0mom34_afterfit}
and \ref{fig:pi0angle_cos34_afterfit} show the distributions of the
reconstructed \pzero\ momentum and direction with and without the
vertex activity after fitting.

The ratio of the NC coherent \pzero\ production to the total CC cross
sections from the MC prediction based on the Rein and Sehgal model is
$1.21~\times10^{-2}$. Hence, the cross section ratios are measured to
be:
\begin{eqnarray}
  \label{eqn:sciboone_nccoh}
  \frac{\sigma({\rm NC coh}\pi^{\rm{0}})}{\sigma({\rm{CC}})}
  &=& {\rm R}_{\rm coh} \times  \frac{\sigma({\rm NC coh}\pi^{\rm{0}})_{MC}}{\sigma({\rm{CC}})_{MC}},\nonumber \\
  &=& {\rm R}_{\rm coh} \times 1.21~\times10^{-2},  \nonumber \\
  &=& (1.16 \pm 0.24) \times 10^{-2},
\end{eqnarray}
where ${\rm R}_{\rm coh}$ is 0.96$\pm$0.20. The mean neutrino energy
for NC coherent \pzero\ events in the sample is estimated\footnote{In
the previous paper \cite{Kurimoto:2009wq}, the mean neutrino energy
was 1.0 GeV despite using the same event sample as this paper. This is
due to a different definition of average neutrino energy. In the
previous paper, we used mean neutrino energy of all events passing the
selection cuts in the MC simulation while, in this paper, we divide
the selected neutrino energy distribution by the coherent cross
section for each neutrino energy bin before calculating the average of
the distribution. The latter method matches SciBooNE's CC coherent
result~\cite{Hiraide:2008eu}} to be 0.8 GeV.  The fractional error of
this cross section ratio is 21\% while the previous result's
fractional error is 60\% ($(0.68\pm0.41)\times10^{-2}$). Hence, the
result has been improved by a factor of three with the new analysis
using vertex activity.  This result is 5.8 standard deviations above
the no coherent production assumption.  The measured cross section is
also consistent with the MC prediction based on the Rein and Sehgal
model~\cite{Rein:1982pf}.  The result is evidence of non-zero coherent
pion production via neutral current interactions at mean neutrino
energy 0.8 GeV.

\section{Discussion}
\label{sec:discussion}

\subsection{Comparison with the CC measurement}

The SciBooNE collaboration measured the ratio of the CC coherent pion
to total CC production as
\begin{eqnarray}
  \label{eqn:sciboone_cccoh}
  \frac{\sigma({\rm CC coh}\pi^{\rm +})}{\sigma({\rm CC})} && \nonumber \\
  &=& (0.16\pm0.17({\rm stat})^{+0.30}_{-0.27}({\rm sys}))\times10^{-2},
\end{eqnarray}
at 1.1 GeV \cite{Hiraide:2008eu}. 
According to Eq.~\ref{eqn:sciboone_nccoh} and \ref{eqn:sciboone_cccoh}, 
the ratio of CC coherent pion production to NC coherent production 
is measured to be 
\begin{eqnarray}
  \frac{\sigma({\rm CC coh}\pi^{\rm +})}{\sigma({\rm NC coh}\pi^{\rm 0})} 
  &=&  \frac{\sigma({\rm CC coh}\pi^{\rm +})}{\sigma({\rm CC})}/
  \frac{\sigma({\rm NC coh}\pi^{\rm 0})}{\sigma({\rm CC})}, \nonumber \\ 
  &=& 0.14^{+0.30}_{-0.28}.
\end{eqnarray}
In contrast, the Rein and Sehgal model \cite{Rein:1982pf} as well as
many other models predict $\sigma({\rm CC coh}\pi^{\rm +})/\sigma({\rm
NC coh}\pi^{\rm 0})=2$ without the lepton mass correction
\cite{Boyd:2009zz}.  Even if we take it into account that the neutrino
energy of the CC measurement (1.1 GeV) is higher than that of NC
measurement (0.8 GeV), the corrected ratio ends up with a smaller
value because the cross section increases with neutrino energy.  So
far, there is no model which can accommodate our measurement of the
CC/NC coherent pion production ratio at these energies, although the
measurement of the ratio at higher energies (\cite{Grabosch:1985mt},
$\sim$7 GeV) is consistent with 2.

\subsection{Comparison with the MiniBooNE measurement}
The MiniBooNE collaboration measured the ratio of NC coherent pion to
NC single \pzero\ production to be:
\begin{eqnarray}
  \label{eqn:miniboone_nccoh}
  \frac{\sigma({\rm NC coh}\pi^{\rm 0})}{\sigma({\rm NC coh}\pi^{\rm 0})+\sigma({\rm NC res}\pi^{\rm 0})} \nonumber \\ 
  = (19.5\pm1.1({\rm stat})\pm2.5({\rm sys}))\%,
\end{eqnarray}
below 2.0 GeV \cite{AguilarArevalo:2008xs}, where 
$\sigma({\rm NC coh}\pi^{\rm 0})$ is the cross section for coherent \pzero\
production and $\sigma({\rm NC res}\pi^{\rm 0})$ is the cross
section for exclusive NC resonant single \pzero\ production. The
qualifier ``exclusive'' in the latter cross section definition by
the MiniBooNE collaboration refers to a neutrino interaction
produced in the resonant channel and with a single \pzero\ in the
final state.  Using our fit result (${\rm R}_{\rm coh}$, 
${\rm R}_{\rm inc}$) shown in Eq.~\ref{eqn:fitresult}, for SciBooNE, the
ratio of NC coherent pion to NC single \pzero\ production is found
to be:
\begin{eqnarray}
  \label{eqn:sciboone_nccoh2}
  \frac{{\rm R}_{\rm coh}\times\sigma({\rm NC coh}\pi^{\rm 0})_{\rm MC}}
       {{\rm R}_{\rm coh}\times\sigma({\rm NC coh}\pi^{\rm 0})_{\rm MC}+{\rm R}_{\rm inc}\times\sigma({\rm NC res}\pi^{\rm 0})_{\rm MC}} \nonumber \\
       = (17.9\pm4.1)\%,
\end{eqnarray}
where we assume that ${\rm R}_{\rm inc}$ scales the NC single resonant
\pzero\ production (although ${\rm R}_{\rm inc}$ actually scales all
incoherent \pzero\ production including the multi meson
production). In fact, single resonant \pzero\ production is dominates
the incoherent \pzero\ sample, comprising 81\% after event selections.
According to Eq.~\ref{eqn:miniboone_nccoh} and
\ref{eqn:sciboone_nccoh2}, the SciBooNE measurement agrees with the
MiniBooNE result within uncertainties.  It should be noted that
MiniBooNE uses a ${\rm C}{\rm H}_2$ target and includes diffractive
hydrogen scattering in their simulation while SciBooNE uses a ${\rm
C}{\rm H}$ target and does not include diffractive hydrogen scattering
in the simulation. However, the effect of these differences is less
than 10\%, which is much smaller than the uncertainty of the SciBooNE
measurement (23\%).

\section{Conclusion}

In conclusion, we have observed NC coherent \pzero\ production at mean
neutrino energy 0.8 GeV.  The ratio of the NC coherent \pzero\
production to the total CC cross sections is measured to be
$1.16~\times10^{-2}$ based on the Rein and Sehgal model.  Our
measurement confirms the previous MiniBooNE result.  The ratio of CC
coherent $\pi^{\rm +}$ to NC coherent \pzero\ production is calculated
to be $0.14^{+0.30}_{-0.28}$ using SciBooNE's previous CC coherent
pion measurement while many models predict 2 as this ratio. We know of
no model that can accommodate our measurement of the CC/NC coherent
pion production ratio.

\section{Acknowledgements}
\label{sec:acknowledgments}

We acknowledge the Physics Department at Chonnam National University,
Dongshin University, and Seoul National University for the loan of
parts used in SciBar and the help in the assembly of SciBar.  We wish
to thank the Physics Departments at the University of Rochester and
Kansas State University for the loan of Hamamatsu PMTs used in the
MRD.  We gratefully acknowledge support from Fermilab as well as
various grants, contracts and fellowships from the MEXT and JSPS
(Japan), the INFN (Italy), the Ministry of Science and Innovation and
CSIC (Spain), the STFC (UK), and the DOE and NSF (USA).  This work was
supported by MEXT and JSPS with the Grant-in-Aid for Scientific
Research A 19204026, Young Scientists S 20674004, Young Scientists B
18740145, Scientific Research on Priority Areas ``New Developments of
Flavor Physics'', and the global COE program ``The Next Generation of
Physics, Spun from Universality and Emergence''.  The project was
supported by the Japan/U.S. Cooperation Program in the field of High
Energy Physics and by JSPS and NSF under the Japan-U.S. Cooperative
Science Program.

\end{document}